\documentstyle[amssymb,12pt]{article}

\begin{document}

\title{HAMILTONIAN EQUATIONS IN~${\mathbb R}^3$}
\author{Ahmet Ay, Metin G{\" u}rses and Kostyantyn Zheltukhin\\
{\small Department of Mathematics, Faculty of Sciences}\\
{\small Bilkent University, 06800 Ankara, Turkey}\\}

\begin{titlepage}
\maketitle

\begin{abstract}

Hamiltonian formulation of $N=3$ systems is considered in general.
The most general solution of the Jacobi equation in ${\mathbb
R}^3$ is proposed. The form of the solution is shown to be valid
also in the neighborhood of some irregular points. Compatible
Poisson structures and corresponding bi-Hamiltonian systems are
also discussed. Hamiltonian structures, classification of
irregular points and the corresponding reduced first order
differential equations of several examples are given.
\end{abstract}

\end{titlepage}

\section*{1. Introduction.}

Hamiltonian formulation of a system of dynamical equations is
important not only in mathematics but also in physics and other
branches of natural sciences. They in general describe conserved
systems. Among all possible odd dimensional cases the three
dimensional dynamical systems have a unique position. The Jacobi
equation in this case reduces to a single scalar equation for
three components of the Poisson structure $J$. Due to this
property $N=3$ dynamical systems attracted many researches to
derive new Hamiltonian systems, \cite{BF}--\cite{GHHFC}. More
recently \cite{ber}, \cite{ber2} a large class of solutions of the
Jacobi equation in ${\mathbb R}^3$ was given. Poisson structures,
in all dimensions, were also considered in \cite{GMP}. In this
work, we consider general solution of Jacobi equation in ${\mathbb
R}^3$. We find the compatible Poisson structures and give the
corresponding bi-Hamiltonian systems. We give all explicit
examples in a special section and a table at the end.

Let us give necessary information about the Poisson structures in
${\mathbb R}^3$. A matrix $J=(J_{ij}), \quad i,j=1,2,3,$ defines a
Poisson structure in ${\mathbb R}^3$ if it is skew-symmetric,
$J_{ij}=-J_{ji}$, and its entries satisfy the Jacobi equation
\begin{equation}\label{jacobi1}
J^{li}\partial_l\,J^{jk}+J^{lj}\partial_l\,J^{ki}+J^{lk}\partial_l\,J^{ij}=0,
\end{equation}
where $i,j,k=1,2,3$. Here we use the summation convention, meaning
that repeated indices are summed up. Let us introduce the
following notations. For matrix $J$ put $J_{12}=u$, $J_{31}=v$,
$J_{23}=w$. Then Jacobi equation (\ref{jacobi1}) takes the form
\begin{equation}\label{jacobi2}
u\partial_1v - v\partial_1u + w\partial_2u -u\partial_2w +
v\partial_3w-w\partial_3v=0.
\end{equation}
It can also be rewritten as
\begin{equation}\label{jacobi3}
u^2\partial_1\frac{v}{u} + w^2\partial_2\frac{u}{w} +
v^2\partial_3\frac{w}{v}=0.
\end{equation}
(We assume that none of the functions $u,v$ and $w$ vanish. If any
one of these functions vanishes then the equation (\ref{jacobi2})
becomes trivial for remaining two variables, see Remark 1.)

We consider the general solution of Jacobi equation equation
(\ref{jacobi3}) and show that it has the following form
\begin{equation}\label{GenSol}
J^{ij}=\mu\epsilon^{ijk}\partial_k\Psi,
\end{equation}
where $\mu$ and $\Psi$ are arbitrary differentiable functions of
$x_{i}$, $i=1,2,3$ and $\epsilon^{ijk}$ is the Levi-Civita symbol.
We also consider special solutions given by

\begin{equation}\label{specialcase}
u\partial_1v - v\partial_1u=0,\; w\partial_2u
-u\partial_2w=0,\quad \textrm {which implies} \quad
v\partial_3w-w\partial_3v=0.
\end{equation}

\noindent Such Poisson structures appear in many examples. We show
that this special class of solutions belongs to the general form
(\ref{GenSol}). We introduce these special solutions to study the
irregular points of the Poisson structures. All the irregular
points of the Poisson structure matrix $J$ given in the examples
\cite{ber}, we know so far, come form this special form. Hence
they are also irregular points of the form (\ref{GenSol}) we give.

\section*{2. The General Solution.}

Assuming that $u \ne 0$, let $\displaystyle{ \rho=\frac{v}{u}}$
and $\displaystyle{ \chi=\frac{w}{u}}$ then equation
(\ref{jacobi2}) can be written as
\begin{equation}\label{jactr1}
\partial_1\rho-\partial_2\chi+\rho\partial_3\chi-\chi\partial_3\rho=0.
\end{equation}
This equation can be put in a more suitable form by writing it
\begin{equation}\label{jactr2}
(\partial_1-\chi\partial_3)\rho-(\partial_2-\rho\partial_3)\chi=0.
\end{equation}
Introducing differential operators $D_1$ and $D_2$ defined by
\begin{equation}
D_1=\partial_{1}-\chi\partial_{3} \qquad
D_2=\partial_{2}-\rho\partial_{3},
\end{equation}
one can write equation (\ref{jactr2}) as
\begin{equation}
D_1\rho-D_2\chi=0. \label{eq1}
\end{equation}

\vspace{0.3cm}

\noindent {\bf Lemma 1.} {\it Let equation (\ref{eq1}) be
satisfied. Then there are new coordinates $\bar{x}_1, \bar{x}_2,
\bar{x}_3$ such that
\begin{equation}\label{nD}
D_1=\partial_{\bar{x}_1},\quad {\mbox and} \quad
D_2=\partial_{\bar{x}_2}.
\end{equation}}

\vspace{0.3cm}

\noindent {\bf Proof.} If equation (\ref{eq1}) is satisfied, it is
easy to show that the operators $D_{1}$ and $D_{2}$ commute, i.e.,
$$
D_1\circ D_2-D_2\circ D_1=0.
$$
Hence, by the Frobenius theorem (see \cite{Olv} p. 40) there exist
coordinates $\bar{x}_1, \bar{x}_2, \bar{x}_3$ such that the
equalities (\ref{nD}) hold. $\Box$

\vspace{0.3cm}

\noindent The coordinates $\bar{x}_1, \bar{x}_2, \bar{x}_3$ are
described by the following lemma.

\vspace{0.3cm}

\noindent {\bf Lemma 2.} {\it Let $\zeta$ be a common invariant
function of $D_1$ and $D_2$, i.e.
\begin{equation}
D_1\zeta=D_2\zeta=0 \label{eq2},
\end{equation}
then the coordinates $\bar{x}_1, \bar{x}_2, \bar{x}_3$ of Lemma 1
are given by
\begin{equation}
\bar{x}_1=x_1,\quad \bar{x}_2=x_2,\quad \bar{x}_3=\zeta.
\end{equation}
Moreover from (\ref{eq2}) we get,
\begin{equation}\label{rhochi}
\chi=\frac{\partial_1\zeta}{\partial_3\zeta},\quad
\rho=\frac{\partial_2\zeta}{\partial_3\zeta}.
\end{equation}
}

\vspace{0.3cm}

\noindent {\bf Theorem 1.} {\it All Poisson structures in
${\mathbb R}^3$, except at some irregular points, take the form
(\ref{GenSol}), i.e., $J_{ij}=\mu\, \epsilon_{ijk}\,
\partial_{k}\, \zeta$. Here $\mu$ and $\zeta$ are some
differentiable functions in ${\mathbb R}^3$ }

\vspace{0.3cm}

\noindent {\bf Proof.} Using (\ref{rhochi}), the entries of matrix
$J$, in the coordinates $\bar{x}_1, \bar{x}_2, \bar{x}_3$, can be
written as
\begin{equation}
\begin{array}{llr}
u&=& \mu\partial_3\zeta,\\
v&=& \mu\partial_2\zeta,\\
w&=& \mu\partial_1\zeta.\\
\end{array}
\end{equation}
Thus matrix $J$ has the form (\ref{GenSol})\, ($\Psi=\zeta$).
$\Box$

\vspace{0.3cm}

\noindent {\bf Remark 1}.\, So far we assumed that $u \ne 0$. If
$u=0$ then the Jacobi equation becomes quite simpler $v
\partial_{3}\, w- w \partial_{3}\, v=0$ which has the simple
solution $w= v \, \xi(x_{1}, x_{2})$, where $\xi$ is an arbitrary
differentiable of $x_{1}$ and $x_{2}$. This class is also covered
by the general solution (\ref{GenSol}) by letting $\Psi$
independent of $x_{3}$.

\vspace{0.3cm}

\noindent A well known example of a dynamical system with Poisson
structure of the form (\ref{GenSol}) is the Euler equations.

\vspace{0.3cm}

\noindent {\bf Example 1.} Consider the Euler equations
(\cite{Olv}, pp.397--398,)
\begin{equation}\label{e}
\begin{array}{lll}
\dot{x}_1&=&\displaystyle{\frac{I_2-I_3}{I_2I_3}}x_2x_3,\\
\dot{x}_2&=&\displaystyle{\frac{I_3-I_1}{I_3I_1}}x_3x_1,\\
\dot{x}_3&=&\displaystyle{\frac{I_1-I_2}{I_1I_2}}x_1x_2,\\
\end{array}
\end{equation}
where $ I_1,I_2,I_3\in {\mathbb R}$ are some (non-vanishing) real
constants. This system admits Hamiltonian representation of the
form~(\ref{GenSol}). The matrix $J$ can be defined in terms of
function $\Psi=-\frac{1}{2}(x^2_1+x^2_2+x^2_3)$ and $\mu=1$, so
\begin{equation}\label{Eulstr}
\begin{array}{lll}
u&=&-x_3,\\
v&=&-x_2,\\
w&=&-x_1,\\
\end{array}
\end{equation}
and
$H=\displaystyle{\frac{x^2_1}{2I_1}+\frac{x^2_2}{2I_2}+\frac{x^2_3}{2I_3}}$.\\

Recently, a large set of solutions of the Jacobi equation
(\ref{jacobi3}) satisfying (\ref{specialcase}) was given in
\cite{ber}. For all such solutions Darboux transformation and
Casimir functionals were obtained, see \cite{ber}.

\vspace{0.3cm}

\noindent {\bf Definition 1.} {\it For every domain $\Omega \in
{\mathbb R}^3$ let ${\mathbf I}_a(\Omega)$ be the set of all
solutions of (\ref{specialcase}) defined in $\Omega$ with $u(x),$
$v(x)$ and $w(x)$ being $C^1(\Omega)$.}

\vspace{0.3cm}

\noindent Following \cite{ber} we have.

\vspace{0.3cm}

\noindent {\bf Proposition 1.} {\it Let
${\eta(x_1,x_2,x_3),\psi_i(x_i),\phi_i(x_i)}, \quad i=1,2,3,$ be
arbitrary differentiable functions defined in $\Omega$. Then the
functions
\begin{equation}\label{cl1}
\begin{array}{rrr}
u(x)&=&\eta(x_1,x_2,x_3)\psi_1(x_1)\psi_2(x_2)\phi_3(x_3)\\
v(x)&=&\eta(x_1,x_2,x_3)\psi_1(x_1)\phi_2(x_2)\psi_3(x_3)\\
w(x)&=&\eta(x_1,x_2,x_3)\phi_1(x_1)\psi_2(x_2)\psi_3(x_3)\\
\end{array}
\end{equation}
define a solution of equation (\ref{specialcase}) belonging to
${\mathbf I}_a(\Omega)$. }

\vspace{0.3cm}

\noindent {\bf Definition 2.} {\it For every domain $\Omega
\in{\mathbb R}^3,$ let ${\mathbf I}_b(\Omega)$ be the set of all
solutions of (\ref{specialcase}) defined in $\Omega$ where one of
the functions $u(x),$ $v(x)$ and $w(x)$ is zero and the others are
not identically zero in $\Omega$.}

\vspace{0.3cm}

\noindent Following \cite{ber} we have.

\vspace{0.3cm}

\noindent {\bf Proposition 2.} {\it Let ${\eta(x_1,x_2,x_3),
\xi_i(x_i)}, \quad i=1,2,3,$ be arbitrary differentiable functions
defined in $\Omega$. Then the functions
\begin{equation}\label{cl2u}
u(x)=0,\quad v(x)=\eta(x_1,x_2,x_3)\xi_2(x_2),\quad
w(x)=\eta(x_1,x_2,x_3)\xi_1(x_1)
\end{equation}
define a solution of equation (\ref{jacobi1}) belonging to
${\mathbf I}_b(\Omega),$ $u=0$. Similar solutions can be given in
the case $v=0$ and the case $w=0$. }

\vspace{0.3cm}

\noindent {\bf Remark 2.} All of the Poisson structures described
in \cite{ber} have the form (\ref{GenSol}). For the Poisson
structure $J$, given by (\ref{cl1}), assume $\psi_1$, $\psi_3$,
and $\psi_3$ to be non vanishing and define
$\mu=\eta(x_1,x_2,x_3)\psi_1(x_1)\psi_2(x_2)\psi_3(x_3)$ and
$$
\Psi=\displaystyle{ \int^{x_{1}}
\frac{\phi_1}{\psi_1}dx_1+\int^{x_{2}}
\frac{\phi_2}{\psi_2}dx_2}+\int^{x_{3}} \frac{\phi_3}{\psi_3}dx_3
$$
then $J$ has form (\ref{GenSol}). For the Poisson structure $J$,
given by (\ref{cl2u}), define $\mu=\eta(x_1,x_2,x_3)$ and $\Psi=
\int^{x_1}\xi_1(x_1) + \int^{x_3}\xi_2(x_2)$ then $J$ has form
(\ref{GenSol}).

\vspace{0.3cm}

Let us give two examples of systems that admit Hamiltonian
representation described by the Proposition 1 and Proposition 2.

\vspace{0.3cm}

\noindent {\bf Example 2.} Consider Lotka-Voltera system
\cite{NG}, \cite{N1}
\begin{equation}
\begin{array}{lll}
\dot{x}_1&=&-abcx_1x_3-bc\mu_0 x_1+cx_1x_2+c\nu x_1\\
\dot{x}_2&=&-a^2bcx_2x_3-abc\mu_0 x_2+x_1x_2\\
\dot{x}_3&=&-abcx_2x_3-abc\nu_0 x_3+bx_1x_3\\
\end{array}
\end{equation}
where $a,b,c,\mu_0,\nu_0\in {\mathbb R}$ are constants.\\
The matrix $J$ is given by
\begin{equation}\label{LotVolstr}
\begin{array}{lll}
u&=&cx_1x_2\\
v&=&-bcx_1x_3\\
w&=&-x_2x_3\\
\end{array}
\end{equation}
and $H=abx_1+x_2-ax_3+\nu_0\ln x_2 -\mu_0\ln x_3$. \\

\vspace{0.3cm}

\noindent {\bf Example 3.} Consider Lorenz system \cite{NG}
\begin{equation}
\begin{array}{lll}
\dot{x}_1&=&\displaystyle{\frac{1}{2}}x_2\\
\dot{x}_2&=&-x_1x_3\\
\dot{x}_3&=&x_1x_2.\\
\end{array}
\end{equation}
The matrix $J$ is given by
\begin{equation}\label{Lorstr}
\begin{array}{lll}
u&=&\displaystyle{\frac{1}{4}}\\
v&=&0\\
w&=&-\displaystyle{\frac{1}{2}}x_1\\
\end{array}
\end{equation}
and $H=x^2_2+x^2_3$.\\

\vspace{0.3cm}

\noindent Many other examples are given in the Section 3.

In the derivation of the general solution, Theorem 1, we assumed
that one of the components of matrix $J$ is different from zero.
In addition our derivation is valid only in a neighborhood of a
regular point of $J$ (matrix $J\ne 0$ at this point). If $p\in
{\mathbb R}^3$ is an irregular point where $u(p)=v(p)=w(p)=0$ it
is not clear whether our solution is valid in a neighborhood of
such a point. Here we shall show that the Poisson structures given
by (\ref{GenSol}) preserve their form in the neighborhood of the
following irregular points.

\vspace{0.3cm}

\noindent {\bf Lemma 3.} {\it The solutions of equation defined in
Proposition 1 and Proposition 2 and written in the form
(\ref{GenSol}) preserve their form in the neighborhood of
the irregular points, lines and planes in ${\mathbb{R}}^3$ defined below \\
{\bf a.} Irregular points. Let $p=(p_1,p_2,p_3)$ be such that
$\phi_1(p_1)=\phi_2(p_2)=\phi_3(p_3)=0$ and $\psi_i(p_i)\ne 0$,
$i=1,2,3$ then $p$ is an irregular point
where the general form (\ref{GenSol}) is preserved .\\
{\bf b.} Irregular lines or irregular plains. Let
$p=(p_1,p_{2},p_{3}) \in {\mathbb R}^3$ be such that
$\eta(p_1,x_2,x_3)=0$ ($\eta(x_{1},p_{2},x_{3})=0$ or
$\eta(x_{1},x_{2},p_{3})=0$) and $\psi_i(p_i)\ne 0$, $i=1,2,3$
then $x_1=p_1$ ($x_{2}=p_{2}$ or $x_{3}=p_{3}$) is an irregular
plane, where the general form (\ref{GenSol}) is preserved. Let
$x_1=p_1$, $x_2=p_2$ be such that $\eta(p_1,p_2,x_3)=0$
($\eta(p_{1},x_{2},p_{3})=0$ or $\eta(x_{1},p_{2},p_{3})=0$) and
$\psi_i(p_i)\ne 0$, $i=1,2,3$ then $x_1=p_1$, $x_2=p_2$
($x_{1}=p_{1},x_{3}=p_{3}$ or $x_{2}=p_{2}, x_{3}=p_{3}$) is an
irregular line, where the general
form (\ref{GenSol}) is preserved.} \\

\vspace{0.3cm}

\noindent {\bf Proof.} The solution given in Proposition 1 and
Proposition 2 solves the following equations (without any
division)
\begin{equation}
\begin{array}{c}
u\partial_1 v -v\partial_1 u=0\\
-u\partial_2 w +w\partial_2 u=0\\
v\partial_3 w - w\partial_3 v=0\\
\end{array}
\end{equation}
The general form (\ref{GenSol}), given in Remark 2, is also
preserved at such points since we can define
$\mu=\eta(x_1,x_2,x_3)\psi_1(x_1)\psi_2(x_2)\psi_3(x_3)$ and
$$
\Psi=\displaystyle{ \int^{x_{1}}
\frac{\phi_1}{\psi_1}dx_1+\int^{x_{2}}
\frac{\phi_2}{\psi_2}dx_2}+\int^{x_{3}} \frac{\phi_3}{\psi_3}dx_3.
$$
or if one of the component of $J$ is zero, assume $u=0$, we define
$\mu=\eta(x_1,x_2,x_3)$ and $\Psi= \int^{x_1}\xi_1(x_1) +
\int^{x_3}\xi_2(x_2).$ $\Box$

\vspace{0.3cm}

\noindent{\bf Example 4.} For the Euler system considered in
Example~1 the Poisson structure, given by (\ref{Eulstr}), has
irregular point $p=(0,0,0)$. The irregular point $p=(0,0,0)$
satisfies the conditions of the Lemma 3, the functions
$\Psi=-\frac{1}{2}(x^2_1+x^2_2+x^2_3)$, $\mu=1$ in terms of which
the Poison structure is given, are well defined in a neighborhood
of $p=(0,0,0)$.

\section*{4. Bi-Hamiltonian system.}

In general the Darboux theorem states that (see \cite{Olv}),
locally, all Poisson structures can be reduced to the standard one
(Poisson structure with constant entries). The above theorem,
Theorem 1, resembles the Darboux theorem for $N=3$. All Poisson
structures, at least locally, can be cast into the form
(\ref{GenSol}). This result is important because the Dorboux
theorem is not suitable for obtaining multi-Hamiltonian systems in
${\mathbb R}^3$, but we will show that our theorem is effective
for this purpose. Writing the Poisson structure in the form
(\ref{GenSol}) allows us to construct bi-Hamiltonian
representations of a given Hamiltonian system.

\vspace{0.3cm}

\noindent {\bf Definition 3.} {\it Two Hamiltonian matrices $J$
and $\tilde{J}$ are compatible, if the sum $J+\tilde{J}$ defines
also a Poisson structure.}

\vspace{0.3cm}

\noindent The compatible Poisson structures can be used to
construct bi-Hamiltonian equations.

\vspace{0.3cm}

\noindent {\bf Definition 4.} {\it A Hamiltonian equation is said
to be bi-Hamiltonian if it admits two Hamiltonian representations
with compatible Poisson structures
\begin{equation}
\frac{dx}{dt}=J\nabla H=\tilde{J}\nabla\tilde{H}
\end{equation}
where $J$ and $\tilde{J}$ are compatible.}

\vspace{0.3cm}

\noindent {\bf Lemma 4.} {\it Let Poisson structures $J$ and
$\tilde{J}$ have form (\ref{GenSol}), so
$J^{ij}=\mu\epsilon^{ijk}\partial_k\Psi$ and
$\tilde{J}^{ij}=\tilde{\mu}\epsilon^{ijk}\partial_k\tilde{\Psi}$.
Then $J$ and $\tilde{J}$ are compatible if and only if there exist
a differentiable function $\Phi(\Psi,\tilde{\Psi})$ such that
\begin{equation}
\tilde{\mu}=\mu\frac{\partial_{\tilde{\Psi}}\Phi}{\partial_{\Psi}\Phi},
\label{con}
\end{equation}
provided that $\partial_{\Psi}\, \Phi \equiv {\partial \Phi \over
\partial \Psi} \ne 0$ and $\partial_{\tilde{\Psi}}\, \Phi \equiv
{\partial \Phi \over \partial \tilde{\Psi}} \ne 0$ }

\vspace{0.3cm}

\noindent This suggests that all Poisson structures in ${\mathbb
R}^3$ have compatible pairs, because the condition (\ref{con}) is
not so restrictive on the Poisson matrices $J$ and ${\tilde J}$.
Such compatible Poisson structures can be used to construct
bi-Hamiltonian systems.

\vspace{0.3cm}

\noindent {\bf Lemma 5.} {\it Let $J$ be given by (\ref{GenSol})
and $H(x_1,x_2,x_3)$ is any differentiable function then the
Hamiltonian equation
\begin{equation}
\frac{dx}{dt}=J\nabla H =- \mu\, {\nabla \Psi} \times {\nabla H},
\end{equation}
is bi-Hamiltonian with the second structure given by $\tilde{J}$
with entries
\begin{equation}
\begin{array}{ccc}
\tilde{u}(x)&=&\tilde{\mu}\, \partial_1g(\Psi(x_1x_2x_3),H(x_1,x_2,x_3)),\\
\tilde{v}(x)&=&-\tilde{\mu}\, \partial_2g(\Psi(x_1x_2x_3),H(x_1,x_2,x_3)),\\
\tilde{w}(x)&=&\tilde{\mu}\, \partial_3g(\Psi(x_1x_2x_3),H(x_1,x_2,x_3)),\\
\end{array}
\end{equation}
and $\tilde{H}=h(\Psi(x_1x_2x_3),H(x_1,x_2,x_3))$,
$\tilde{\Psi}=g(\Psi(x_{1},x_{2},x_{3}), H (x_{1},x_{2},x_{3}))$,
$\tilde{\mu}=\mu\frac{\partial_{\tilde{\Psi}}\Phi}{\partial_{\Psi}\Phi}$.
Provided that there exist differentiable functions $\Phi
(\Psi,{\tilde{\Psi}})$, $h(\Psi,H)$, and $g(\Psi,H)$ satisfying
the following equation
\begin{equation}\label{f2}
\frac{\partial g}{\partial \Psi} \frac{\partial h}{\partial H}
-\frac{\partial g}{\partial H}\frac{\partial h}{\partial
\Psi}=\frac{\Phi_1(\Psi,g)}{\Phi_2(\Psi,g)},
\end{equation}
where $\Phi_1=\partial_\Psi\Phi|_{(\Psi,g)}$,
$\Phi_2=\partial_{\tilde{\Psi}}\Phi|_{(\Psi,g)}$.}

\vspace{0.3cm}

\noindent{\bf Proof.} By Lemma~4, $J$ and $\tilde{J}$ are
compatible and it can be shown by a straightforward calculation
that the equality (being a bi-Hamiltonian system)
\begin{equation}
\tilde{J}\nabla \tilde{H}=J\nabla H
\end{equation}
or
\begin{equation}
\tilde{\mu}\, \nabla\, \tilde{\Psi} \times \nabla \tilde{H}= \mu\,
\nabla \, \Psi \times \nabla\, H
\end{equation}

\noindent is guaranteed by (\ref{f2}). Hence the system
\begin{equation}
\begin{array}{lll}\label{sys}
\displaystyle{\frac{dx_1}{dt}}&=&\mu\partial_3\Psi\partial_2H
-\partial_2\Psi\partial_3H\\
\displaystyle{\frac{dx_2}{dt}}&=&-\mu\partial_3\Psi\partial_1H
+\partial_1\Psi\partial_3H\\
\displaystyle{\frac{dx_3}{dt}}&=&\mu\partial_2\Psi\partial_1H
-\partial_1\Psi\partial_2H\\
\end{array}
\end{equation}
is bi-Hamiltonian. $\Box$

\vspace{0.3cm}

\noindent {\bf Remark 3}.\, The Hamiltonian function $H$ is a
conserved quantity of the system. It is clear from the expression
(\ref{sys}) that the function $\Psi$ is another conserved quantity
of the system. Hence for a given Hamiltonian system there is a
duality between $H$ and $\Psi$. Such a duality arises naturally
because a simple solution of the equation (\ref{f2}) is
$\tilde{\Psi}=H$, $\tilde{H}=\Psi$ and $\tilde{\mu}=-\mu$. Thus we
have an hierarchy of Hamiltonians that starts with a Casimir of
the second structure and terminates with a Casimir of the first
structure. Such systems are equivalent to the quasi-bi-Hamiltonian
systems of lower dimension with non-degenerate Poisson structures
(see \cite{Blz} pp. 185-220).

\vspace{0.3cm}

\noindent {\bf Remark 4}.\, Using the Lemma~5 we can construct
infinitely many compatible Hamiltonian representations by choosing
functions $\Phi,\, g,\, h$ satisfying (\ref{f2}). If we fix
functions $\Phi$, and $g$ then equation (\ref{f2}) became linear
first order partial differential equations for $h$. For instance,
takeing $g=\Psi H$ and $\tilde{\mu}=-\mu$, which fixes $\Phi$, we
obtain $h=\ln H$. Thus we a obtain second Hamiltonian
representation with $\tilde{J}$ given by $\tilde{\Psi}=\Psi H$ and
$\tilde{H}=\ln H$.

\section*{4. Examples.}

\noindent Let us give examples of Hamiltonian systems. For each
Hamiltonian system we give the Hamiltonian $H$ and functions
$\Psi$ and $\mu$ in terms of which the corresponding Poisson
structure may be written, using (\ref{GenSol}). Functions $H$ and
$\Psi$ are first integrals of the system so one can use them to
reduce the system to a first order ordinary differential equation.
We give the reduced equation for the examples. We also give
irregular points for the Poison structures. For all examples
except Example~7 the form of the Poisson structure (\ref{GenSol})
is preserved in a neighborhood of irregular points (function
$\Psi$ and $\mu$ are well defined). For Example~7 the form of the
Poisson structure (\ref{GenSol}) is not preserved, the function
$\Psi$ is not defined in a neighborhood of irregular points but
the Hamiltonian function is also not defined at the irregular
points. Hence this system dose not have a Hamiltonian formulation
in the neighborhood of such points. Examples 6-12 satisfy the
special case given in Proposition 1 and Proposition 2. Please see
\cite{ber} for the examples and related references.

\vspace{0.3cm}

\noindent {\bf Example 6.} For the Euler system considered in
Example~1 we gave Poisson structure in terms of functions
$\Psi,\mu$ and the Hamiltonian. The reduced equations are
\begin{equation}
\begin{array}{lll}
x_1&=&(C_1 +\frac{I_1(I_3-I_2)}{I_3(I_2-I_1)}x_3^2)^{\frac{1}{2}} \\
x_2&=&(C_2 +\frac{I_2(I_3-I_1)}{I_3(I_1-I_2)}x_3^2)^{\frac{1}{2}} \\
\dot{x}_3&=&(C_1
+\frac{I_1(I_3-I_2)}{I_3(I_2-I_1)}x_3^2)^{\frac{1}{2}}
(C_2 +\frac{I_2(I_3-I_1)}{I_3(I_1-I_2)}x_3^2)^{\frac{1}{2}} \\
\end{array}
\end{equation}
The Poisson structure is given by (\ref{Eulstr}). It has an
irregular point $p=(0,0,0)$ (the origin).

\vspace{0.3cm}

\noindent {\bf Example 7.} The Lotka-Voltera system considered in
in Example~2 has the matrix $J$ given by $\Psi=-\ln x_1-b\ln
x_2+c\ln x_3$, $\mu=x_1x_2x_3$ and
the Hamiltonian $H=abx_1+x_2 -ax_3+\nu_0\ln x_2-\mu_0\ln x_3$. \\
The reduced equations can be obtained using equalities
\begin{equation}
\begin{array}{lll}
-\ln x_1-b\ln x_2+c\ln x_3&=&C_1\\
abx_1+x_2 -ax_3+\nu_0\ln x_2-\mu_0\ln x_3&=&C_2\\
\end{array}
\end{equation}
The Poisson structure is given by (\ref{LotVolstr}). It has
irregular lines given by $x_i=0$ and $x_j=0$ , $i,j=1,2,3$, $j\ne
i$ (coordinate lines). Both $\Psi$ and $H$ are not defined at
these points. So, the system dose not have a Hamiltonian
formulation at these points.

\vspace{0.3cm}

\noindent {\bf Example 8.} The Lorentz system considered in
Example~3 has the matrix $J$ given by
$\Psi=\displaystyle{\frac{1}{4}}(x_3-x_1^2)$, $\mu=1$ and
the Hamiltonian $H=x_1^2+x_3^2$. \\
The reduced equations are
\begin{equation}
\begin{array}{lll}
x_1&=&(C_1-x_3)^{\frac{1}{2}}\\
x_2&=&(C_2-x_3^2)^{\frac{1}{2}}\\
\dot{x}_3&=&(C_1-x_3)^{\frac{1}{2}}(C_2-x_3^2)^{\frac{1}{2}}\\
\end{array}
\end{equation}
The Poisson structure is given by (\ref{Lorstr}). It has no
irregular points.

\vspace{0.3cm}

\noindent {\bf Example 9.} Consider Kermac-Mackendric system
\cite{NG}, \cite{N2}
\begin{equation}
\begin{array}{lll}
\dot{x}_1&=&-rx_1x_2\\
\dot{x}_2&=&rx_1x_2-ax_2\\
\dot{x}_3&=&ax_2\\
\end{array}
\end{equation}
where $r,a\in {\mathbb R}$ are constants.\\
The matrix $J$ is given by $\Psi=x_1+x_2+x_3$, $\mu=x_1x_2$ and
the Hamiltonian is $H=rx_3+a\ln x_1$. \\
The reduced equations are
\begin{equation}
\begin{array}{lll}
x_2&=&C_1+\frac{a}{r}\ln x_1 -x_1\\
x_3&=&C_2-\frac{a}{r}\ln x_1\\
\dot{x}_1&=&-rx_1(C_1+\frac{a}{r}\ln x_1 -x_1)\\
\end{array}
\end{equation}
The Poisson structure is given by
\begin{equation}
\begin{array}{lll}
u&=&x_1x_2\\
v&=&x_1x_2\\
w&=&x_1x_2.\\
\end{array}
\end{equation}
It has irregular planes $x_1=0$ and $x_2=0$ (coordinate planes).

\vspace{0.3cm}

\noindent {\bf Example 10.} Consider May-Leonard system \cite{NG}
\begin{equation}
\begin{array}{lll}
\dot{x}_1&=&-x_2^{-\alpha} x_3^{-\alpha}\\
\dot{x}_2&=&-x_1^{-\alpha} x_3^{-\alpha}\\
\dot{x}_3&=&-x_1^{-\alpha} x_2^{-\alpha}.\\
\end{array}
\end{equation}
The matrix $J$ is given by
$\Psi=\displaystyle{\frac{1}{(1-\alpha)^2}}(x_2^{1-\alpha}-x_1^{1-\alpha})$,
$\mu=1$ and the Hamiltonian is $H=x_1^{1-\alpha}-x_3^{1-\alpha}$, $\alpha<0$.\\
The reduced equations are
\begin{equation}
\begin{array}{lll}
x_2&=&(C_1+x_1^{1-\alpha})^{\frac{1}{1-\alpha}}\\
x_3&=&(C_2+x_1^{1-\alpha})^{\frac{1}{1-\alpha}}\\
\dot{x}_1&=&-(C_1+x_1^{1-\alpha})^{\frac{\alpha}{1-\alpha}}
(C_2+x_1^{1-\alpha})^{\frac{\alpha}{1-\alpha}}\\
\end{array}
\end{equation}
The Poisson structure $J$ is given by
\begin{equation}
\begin{array}{lll}
u&=&0\\
v&=&\displaystyle{\frac{x_2^{-\alpha}}{\alpha -1}}\\
w&=&\displaystyle{\frac{x_1^{-\alpha}}{\alpha -1}}\\
\end{array}
\end{equation}
It has an irregular line $x_1=0$, $x_2=0$ (coordinate line).

\vspace{0.3cm}

\noindent {\bf Example 11.} Consider Maxvel-Bloch system \cite{NG}
\begin{equation}
\begin{array}{lll}
\dot{x}_1&=&x_2\\
\dot{x}_2&=&x_1x_3\\
\dot{x}_3&=&-x_1x_2.\\
\end{array}
\end{equation}
The matrix $J$ is given by
$\Psi=-\displaystyle{\frac{1}{2\upsilon}}(x_2^2+x_3^2)$, $\mu=1$
and the Hamiltonian is
$H=\displaystyle{\frac{1}{2}\alpha(x_2^2+x_3^2)- \frac{1}{\upsilon}(x_3+x_1^2)}$, $\upsilon\ne 0$.\\
The reduced equations are
\begin{equation}
\begin{array}{lll}
x_1&=&(C_1+\frac{\alpha v}{2}C_2-x_3)^{\frac{1}{2}}\\
x_2&=&(C_2-x_3^2)^{\frac{1}{2}}\\
\dot{x}_3&=&-(C_1+\frac{\alpha v}{2}C_2-x_3)^{\frac{1}{2}}(C_2-x_3^2)^{\frac{1}{2}}\\
\end{array}
\end{equation}
The Poisson structure is given by
\begin{equation}
\begin{array}{lll}
u&=&\displaystyle{\frac{-1}{\upsilon}}x_3\\
v&=& \displaystyle{\frac{-1}{\upsilon}}x_2\\
w&=&0\\
\end{array}
\end{equation}
It has an irregular line $x_2=0$, $x_3=0$ (coordinate line).

\vspace{0.3cm}

\noindent {\bf Example 12.} Consider systems that are obtained
from Lorenz system~\cite{L}
\begin{equation}
\begin{array}{lll}
\dot{x}&=&\sigma(x-y)\\
\dot{y}&=&-y+rx-xz\\
\dot{z}&=&-bz+xy.\\
\end{array}
\end{equation}
Following \cite{GHHFC}, for appropriate subset of parameters by recalling we have:\\
{\bf Lorentz(1) system}
\begin{equation}
\begin{array}{lll}
\dot{x}_1&=&\sigma x_2e^{(\sigma-1)t}\\
\dot{x}_2&=&x_1e^{(1-\sigma)t}(r-x_3e^{-2\sigma t})\\
\dot{x}_3&=&x_1x_2e^{(\sigma-1)t}.\\
\end{array}
\end{equation}
The matrix $J$ is given by
$\Psi=-\displaystyle{\frac{r}{4\sigma}}x_1^2e^{(1-\sigma)t}-
\displaystyle{\frac{1}{4}}x_2^2e^{(\sigma-1)t}-
\displaystyle{\frac{1}{4}}x_3^2e^{(1-3\sigma)t}$, $\mu=1$ and
the Hamiltonian is $H=x^2_1-2\sigma x_3$.\\
The reduced equations are
\begin{equation}
\begin{array}{lll}
x_1&=&(C_1+2\sigma x_3)^{\frac{1}{2}}\\
x_2&=&(C_2-\frac{r}{\sigma}(C_1+2\sigma x_3)e^{2(1-\sigma)t}
-x_3^2e^{2(1-2\sigma)t})^{\frac{1}{2}}\\
\dot{x}_3&=&(C_1+2\sigma
x_3)^{\frac{1}{2}}(C_2-\frac{r}{\sigma}(C_1+2\sigma
x_3)e^{2(1-\sigma)t} -x_3^2e^{2(1-2\sigma)t})^{\frac{1}{2}}e^{(1-\sigma)t}\\
\end{array}
\end{equation}
The Poisson structure is given by
\begin{equation}
\begin{array}{lll}
u&=&\displaystyle{\frac{1}{2}}x_3e^{(1-3\sigma)t}\\
v&=&\displaystyle{\frac{1}{2}}x_2e^{(\sigma-1)t}\\
w&=&\displaystyle{-\frac{r}{2\sigma}}x_1e^{(1-\sigma)t}\\
\end{array}
\end{equation}
It has an irregular point $x_1=0$, $x_2=0$, $x_3=0$ (the origin).\\
{\bf Lorentz(3) system}
\begin{equation}
\begin{array}{lll}
\dot{x}_1&=&\sigma x_2e^{(\sigma-1)t}\\
\dot{x}_2&=&-x_1x_3e^{-\sigma t}\\
\dot{x}_3&=&x_1x_2e^{-\sigma t}.\\
\end{array}
\end{equation}
The matrix $J$ is given by
$\Psi=-\displaystyle{\frac{1}{4}}x_1^2e^{-\sigma t}+
\displaystyle{\frac{\sigma}{2}}x_3e^{(\sigma-1)t}$, $\mu=1$ and
the Hamiltonian is $H=x^2_2+x_3^2$.\\
The reduced equations are
\begin{equation}
\begin{array}{lll}
x_1&=&(C_1e^{\sigma t}+2\sigma x_3e^{(2\sigma-1)t})^{\frac{1}{2}}\\
x_2&=&(C_2-x_3^2)^{\frac{1}{2}}\\
\dot{x}_3&=&(C_1e^{\sigma t}+2\sigma
x_3e^{(2\sigma-1)t})^{\frac{1}{2}}
(C_2-x_3^2)^{\frac{1}{2}}e^{-\sigma t}\\
\end{array}
\end{equation}
The Poisson structure is given by
\begin{equation}
\begin{array}{lll}
u&=&\displaystyle{\frac{1}{2}}\sigma e^{(\sigma-1)t}\\
v&=&0\\
w&=&\displaystyle{-\frac{1}{2}}x_1e^{-\sigma t}\\
\end{array}
\end{equation}
It has no irregular points.\\
{\bf Lorentz(5) system}
\begin{equation}
\begin{array}{lll}
\dot{x}_1&=&x_2\\
\dot{x}_2&=&rx_1-x_1x_3e^{-t}\\
\dot{x}_3&=&x_1x_2e^{-t}.\\
\end{array}
\end{equation}
The matrix $J$ is given by
$\Psi=\displaystyle{\frac{1}{4}}x_1^2e^{-t}-\frac{1}{2}x_3$,
$\mu=1$
and the Hamiltonian is $H=-rx_1^2+x^2_2+x_3^2$.\\
The reduced equations are
\begin{equation}
\begin{array}{lll}
x_1&=&(C_1e^t+2x_3e^t)^{\frac{1}{2}}\\
x_2&=&(C_2+rC_1e^t+2rx_3e^t-x_3^2)^{\frac{1}{2}}\\
\dot{x}_3&=&(C_1e^t+2x_3e^t)^{\frac{1}{2}}
(C_2+rC_1e^t+2rx_3e^t-x_3^2)^{\frac{1}{2}}e^{-t}\\
\end{array}
\end{equation}
The Poisson structure is given by
\begin{equation}
\begin{array}{lll}
u&=& \displaystyle{\frac{1}{2}}\\
v&=&0\\
w&=&\displaystyle{-\frac{1}{2}}x_1e^{-t}\\
\end{array}
\end{equation}
It has no irregular points.

\vspace{0.3cm}

\noindent {\bf Example 13.} Consider systems that are obtained
from Rabinovich system~\cite{PR}
\begin{equation}
\begin{array}{lll}
\dot{x}&=&-\nu_1 x+hy+yz\\
\dot{y}&=&hx-\nu_2y-xz\\
\dot{z}&=&-\nu_3z+xy.\\
\end{array}
\end{equation}
Following \cite{GHHFC}, for appropriate subset of parameters by recalling we have: \\
{\bf Rabinovich (1) system}
\begin{equation}
\begin{array}{lll}
\dot{x}_1&=&hx_2+x_2x_3e^{-2\nu t}\\
\dot{x}_2&=&hx_1-x_1x_3e^{-2\nu t}\\
\dot{x}_3&=&x_1x_2.\\
\end{array}
\end{equation}
The matrix $J$ is given by $\Psi=\displaystyle{\frac{1}{8}}x_1^2 -
\displaystyle{\frac{1}{8}}x_2^2
-\displaystyle{\frac{1}{4}}x_3^2e^{-2\nu t}$, $\mu=1$
and the Hamiltonian is $H=x^2_1+x_2^2-4hx_3$.\\
The reduced equations are
\begin{equation}
\begin{array}{lll}
x_1&=&(C_1+x_3^2e^{-2\nu t}+2hx_3)^{\frac{1}{2}}\\
x_2&=&(C_2-x_3^2e^{-2\nu t}+2hx_3)^{\frac{1}{2}}\\
\dot{x}_3&=&(C_1+x_3^2e^{-2\nu t}+2hx_3)^{\frac{1}{2}}
(C_2-x_3^2e^{-2\nu t}+2hx_3)^{\frac{1}{2}}\\
\end{array}
\end{equation}
The Poisson structure is given by
\begin{equation}
\begin{array}{lll}
u&=&\displaystyle{\frac{1}{2}}x_3e^{-2\nu t}\\
v&=&\displaystyle{\frac{1}{4}}x_2\\
w&=&\displaystyle{-\frac{1}{4}}x_1\\
\end{array}
\end{equation}
It has an irregular point $x_1=0$, $x_2=0$, $x_3=0$ (the origin).\\
{\bf Rabinovich (2) system}
\begin{equation}
\begin{array}{lll}
\dot{x}_1&=&hx_2+x_2x_3e^{-\nu t}\\
\dot{x}_2&=&hx_1-x_1x_3e^{-\nu t}\\
\dot{x}_3&=&x_1x_2e^{-\nu t}.\\
\end{array}
\end{equation}
The matrix $J$ is given by
$\Psi=\displaystyle{\frac{1}{8}}x_1^2e^{-\nu t} +
\displaystyle{\frac{1}{8}} x_2^2e^{-\nu
t}-\displaystyle{\frac{1}{2}}hx_3$, $\mu=1$ and the Hamiltonian is
$H=x^2_1-x_2^2-2x_3^2$.\\
The reduced equations are
\begin{equation}
\begin{array}{lll}
x_1&=&(C_1e^{\nu t}+C_2+x_3^2+2hx_3e^{\nu t})^{\frac{1}{2}}\\
x_2&=&(C_1e^{\nu t}-C_2-x_3^2+2hx_3e^{\nu t})^{\frac{1}{2}}\\
\dot{x}_3&=&(C_1e^{\nu t}+C_2+x_3^2+2hx_3e^{\nu t})^{\frac{1}{2}}
(C_1e^{\nu t}-C_2-x_3^2+2hx_3e^{\nu t})^{\frac{1}{2}}e^{-\nu t}\\
\end{array}
\end{equation}
The Poisson structure is given by
\begin{equation}
\begin{array}{lll}
u&=&\displaystyle{-\frac{1}{2}}h\\
v&=&\displaystyle{\frac{1}{4}}x_2e^{-\nu t}\\
w&=&\displaystyle{\frac{1}{4}}x_1e^{-\nu t}\\
\end{array}
\end{equation}
It has no irregular points.\\
{\bf Rabinovich (3) system}
\begin{equation}
\begin{array}{lll}
\dot{x}_1&=&x_2x_3e^{\nu_3 t}\\
\dot{x}_2&=&-x_1x_3e^{-\nu_3 t}\\
\dot{x}_3&=&x_1x_2e^{(\nu_3-2\nu)t}.\\
\end{array}
\end{equation}
The matrix $J$ is given by $\Psi=
\displaystyle{\frac{1}{4}}x_2^2e^{(\nu_3-2\nu)t}+\displaystyle{\frac{1}{4}}x_3^2e^{-\nu_3
t}$, $\mu=1$
and the Hamiltonian is $H=x_1^2+x_2^2$.\\
The reduced equations are
\begin{equation}
\begin{array}{lll}
x_1&=&(C_1-x_2^2)^{\frac{1}{2}}\\
x_3&=&(C_2e^{-\nu_3 t}-x_3^2e^{-2(\nu-\nu_3)t})^{\frac{1}{2}}\\
\dot{x}_2&=&(C_1-x_2^2)^{\frac{1}{2}}
(C_2e^{-\nu_3 t}-x_3^2e^{-2(\nu-\nu_3)t})^{\frac{1}{2}}e^{(\nu_3-2\nu)t}\\
\end{array}
\end{equation}
The Poisson structure is given by
\begin{equation}
\begin{array}{lll}
u&=&\displaystyle{\frac{1}{2}}x_3e^{-\nu_3 t}\\
v&=&\displaystyle{\frac{1}{2}}x_2e^{(\nu_3-2\nu) t}\\
w&=&0\\
\end{array}
\end{equation}
It has an irregular line $x_2=0$, $x_3=0$ (coordinate line).\\
{\bf Rabinovich (4) system}
\begin{equation}
\begin{array}{lll}
\dot{x}_1&=&hx_2e^{\nu_1 t}+x_2x_3e^{\nu_1 t}\\
\dot{x}_2&=&hx_1e^{-\nu_1 t}-x_1x_3e^{-\nu_1 t}\\
\dot{x}_3&=&x_1x_2e^{-\nu_1 t}.\\
\end{array}
\end{equation}
The matrix $J$ is given by
$\Psi=-\displaystyle{\frac{1}{4}}x_1^2e^{-\nu t} -
\displaystyle{\frac{1}{4}}x_2^2e^{\nu_1 t}+hx_3e^{\nu_1 t}$,
$\mu=1$
and the Hamiltonian is $H=x_2^2+(h-x_3)^2$.\\
The reduced equations are
\begin{equation}
\begin{array}{lll}
x_1&=&(C_1e^{\nu t}-(C_2-(h+x_3))e^{(\nu_1+\nu)t})^{\frac{1}{2}}\\
x_2&=&(C_2-(h-x_3)^2)^{\frac{1}{2}}\\
\dot{x}_3&=&(C_1e^{\nu
t}-(C_2-(h+x_3))e^{(\nu_1+\nu)t})^{\frac{1}{2}}
(C_2-(h-x_3)^2)^{\frac{1}{2}}e^{-\nu_1 t}\\
\end{array}
\end{equation}
The Poisson structure is given by
\begin{equation}
\begin{array}{lll}
u&=&he^{\nu_1 t}\\
v&=&\displaystyle{-\frac{1}{2}}x_2e^{\nu_1 t}\\
w&=&\displaystyle{-\frac{1}{2}}x_1e^{-\nu t}\\
\end{array}
\end{equation}
It has no irregular points.\\
{\bf Rabinovich (5) system}
\begin{equation}
\begin{array}{lll}
\dot{x}_1&=&hx_2e^{-\nu_2 t}+x_2x_3e^{-\nu_2 t}\\
\dot{x}_2&=&hx_1e^{\nu_2 t}-x_1x_3e^{\nu_2 t}\\
\dot{x}_3&=&x_1x_2e^{-\nu_2 t}.\\
\end{array}
\end{equation}
The matrix $J$ is given by
$\Psi=\displaystyle{\frac{1}{4}}x_1^2e^{\nu_2 t} +
\displaystyle{\frac{1}{4}}x_2^2e^{-\nu_2 t}-hx_3e^{\nu_2 t}$,
$\mu=1$
and the Hamiltonian is $H=x_1^2-(h+x_3)^2$.\\
The reduced equations are
\begin{equation}
\begin{array}{lll}
x_1&=&(C_1+(h+x_3)^2)^{\frac{1}{2}}\\
x_2&=&(C_2e^{\nu_2 t}-(C_1+(h-x_3))e^{2\nu_2 t})^{\frac{1}{2}}\\
\dot{x}_3&=&(C_1+(h+x_3)^2)^{\frac{1}{2}}
(C_2-(C_1+(h-x_3))e^{2\nu_2 t})^{\frac{1}{2}}e^{-\nu_2 t}\\
\end{array}
\end{equation}
The Poisson structure is given by
\begin{equation}
\begin{array}{lll}
u&=&-he^{\nu_2 t}\\
v&=&\displaystyle{\frac{1}{2}}x_2e^{-\nu_2 t}\\
w&=&\displaystyle{\frac{1}{2}}x_1e^{\nu_2 t}\\
\end{array}
\end{equation}
It has no irregular points.\\
{\bf Rabinovich (6) system}
\begin{equation}
\begin{array}{lll}
\dot{x}_1&=&x_2x_3e^{(\nu_1-2\nu_3) t}\\
\dot{x}_2&=&-x_1x_3e^{-\nu_1 t}\\
\dot{x}_3&=&x_1x_2e^{-\nu_1 t}.\\
\end{array}
\end{equation}
The matrix $J$ is given by
$\Psi=-\displaystyle{\frac{1}{4}}x_1^2e^{-\nu_1 t} -
\displaystyle{\frac{1}{4}}x_2^2e^{(\nu_1-2\nu_2) t}$, $\mu=1$
and the Hamiltonian is $H=x_2^2+x_3^2$.\\
The reduced equations are
\begin{equation}
\begin{array}{lll}
x_1&=&(C_1e^{\nu_1 t}+x_2^2e^{2(\nu_1-\nu_2) t})^{\frac{1}{2}}\\
x_3&=&(C_2-x_2^2)^{\frac{1}{2}}\\
\dot{x}_2&=&-(C_1e^{\nu_1 t}+x_2^2e^{2(\nu_1-\nu_2)
t})^{\frac{1}{2}}
(C_2-x_2^2)^{\frac{1}{2}}e^{-\nu_1 t}\\
\end{array}
\end{equation}
The Poisson structure is given by
\begin{equation}
\begin{array}{lll}
u&=&0\\
v&=&-\displaystyle{\frac{1}{2}}x_2e^{(\nu_1-2\nu_2) t}\\
w&=&-\displaystyle{\frac{1}{2}}x_1e^{-\nu_1 t}\\
\end{array}
\end{equation}
It has an irregular line $x_1=0$, $x_2=0$ (coordinate line).\\
{\bf Rabinovich (7) system}
\begin{equation}
\begin{array}{lll}
\dot{x}_1&=&x_2x_3e^{-\nu_2 t}\\
\dot{x}_2&=&-x_1x_3e^{(\nu_2-2\nu_3) t}\\
\dot{x}_3&=&x_1x_2e^{-\nu_2 t}.\\
\end{array}
\end{equation}
The matrix $J$ is given by
$\Psi=\displaystyle{\frac{1}{4}}x_1^2e^{(\nu_2-2\nu_3) t} +
\displaystyle{\frac{1}{4}}x_2^2e^{-\nu_2t}$, $\mu=1$
and the Hamiltonian is $H=x_1^2-x_3^2$.\\
The reduced equations are
\begin{equation}
\begin{array}{lll}
x_2&=&(C_1e^{\nu_2 t}-x_1^2e^{2(\nu_2-\nu_3)t})^{\frac{1}{2}}\\
x_3&=&(C_2+x_1^2)^{\frac{1}{2}}\\
\dot{x}_1&=&(C_1e^{\nu_2
t}-x_1^2e^{2(\nu_2-\nu_3)t})^{\frac{1}{2}}
(C_2+x_1^2)^{\frac{1}{2}}e^{-\nu_2 t}\\
\end{array}
\end{equation}
The Poisson structure is given by
\begin{equation}
\begin{array}{lll}
u&=&0\\
v&=&\displaystyle{\frac{1}{2}}x_2e^{\nu_2 t}\\
w&=&\displaystyle{\frac{1}{2}}x_1e^{\nu_2-2\nu_3 t}\\
\end{array}
\end{equation}
It has an irregular line $x_2=0$, $x_3=0$ (coordinate line).

\vspace{0.3cm}

\noindent {\bf Example 14.} Consider systems that are obtained
from RTW system \cite{PR}
\begin{equation}
\begin{array}{lll}
\dot{x}&=&\gamma x+\delta y+z-2y^2\\
\dot{y}&=&\gamma y-\delta x +2xy\\
\dot{z}&=&-2z(x+1).\\
\end{array}
\end{equation}
for appropriate subset of parameters by recalling. Following
\cite{GHHFC} we have:\\
\noindent {\bf RTW(1) system}
\begin{equation}
\begin{array}{lll}
\dot{x}_1&=&\delta x_2 +x_3e^{-2t}-2x_2^2\\
\dot{x}_2&=&-\delta x_1 +2x_1x_2\\
\dot{x}_3&=&-x_1x_3,\\
\end{array}
\end{equation}
where $\delta$ is an arbitrary constant. The matrix $J$ is given
by $\Psi=\displaystyle{\frac{1}{2}}(x_1^2-x_2^2+x_3e^{-t})$,
$\mu=1$ and the Hamiltonian is $H=x_3(2x_2-\delta)$.\\
The reduced equations are
\begin{equation}
\begin{array}{lll}
x_1&=&\left(C_1-x_3e^{-t}+\left(\displaystyle{\frac{C_2+\delta x_3}{2x_3}}\right )^2\right )^{\frac{1}{2}}\\
x_2&=&\displaystyle{\frac{C_2+\delta x_3}{2x_3}}\\
\dot{x}_3&=&-\left (C_1-x_3e^{-t}+\left(\displaystyle{\frac{C_2+\delta x_3}{2x_3}}\right)^2\right)^{\frac{1}{2}}x_3\\
\end{array}
\end{equation}
The Poisson structure is given by
\begin{equation}
\begin{array}{lll}
u&=&\displaystyle{\frac{1}{2}}e^{-2 t}\\
v&=&x_2\\
w&=&x_1\\
\end{array}
\end{equation}
It has no irregular points.\\
{\bf RTW(2) system}
\begin{equation}
\begin{array}{lll}
\dot{x}_1&=&\delta x_2 +x_3e^{-t}-2x_2^2e^{-t}\\
\dot{x}_2&=&-\delta x_1 +2x_1x_2e^{-t}\\
\dot{x}_3&=&-x_1x_3e^{-t},\\
\end{array}
\end{equation}
where $\delta$ is an arbitrary constant. The matrix $J$ is given
by $\Psi=-\displaystyle{\frac{\delta}{2}}(x_1^2+x_2^2)-
x_3x_2e^{-t}$,
$\mu=1$ and the Hamiltonian is $H=x_1^2+x_2^2+x_3$.\\
The reduced equations are
\begin{equation}
\begin{array}{lll}
x_1&=&\left(C_2- x_3-\left(C_1e^{t}-\displaystyle{\frac{\delta }{2}}C_2+\displaystyle{\frac{\delta }{2}}x_3\right)^2\right)^{\frac{1}{2}}\\
x_2&=&C_1e^{t}-\displaystyle{\frac{\delta }{2}}C_2+\displaystyle{\frac{\delta }{2}}x_3\\
\dot{x}_3&=&\left(C_2- x_3-\left(C_1e^{t}-\displaystyle{\frac{\delta }{2}}C_2+\displaystyle{\frac{\delta }{2}}x_3\right)^2\right)^{\frac{1}{2}}x_3e^{-t}\\
\end{array}
\end{equation}
The Poisson structure is given by
\begin{equation}
\begin{array}{lll}
u&=&-x_2e^{-t}\\
v&=&-\delta x_2-x_3e^{-t}\\
w&=&-\delta x_1\\
\end{array}
\end{equation}
It has an irregular point $x_1=0$, $x_2=0$, $x_3=0$ (the origin).\\
{\bf RTW(3) system}
\begin{equation}
\begin{array}{lll}
\dot{x}_1&=&(x_3-2x_2)e^{-t}\\
\dot{x}_2&=&2x_1x_2e^{-t}\\
\dot{x}_3&=&-2x_1x_3e^{-t}.\\
\end{array}
\end{equation}
The matrix $J$ is given by $\Psi=(x_1^2-x_2^2+x_3)e^{-t}$, $\mu=1$
and the Hamiltonian is $H=x_2x_3$.\\
The reduced equations are
\begin{equation}
\begin{array}{lll}
x_1&=&\left(C_1e^{t}- x_3-\displaystyle{\frac{C_2^2}{x_3^2}}\right)^{\frac{1}{2}}\\
x_2&=&\displaystyle{\frac{C_2}{x_3}}\\
\dot{x}_3&=&-2\left(C_1e^{t}- x_3-\displaystyle{\frac{C_2^2}{x_3^2}}\right)^{\frac{1}{2}}x_3e^{-t}\\
\end{array}
\end{equation}
The Poisson structure is given by
\begin{equation}
\begin{array}{lll}
u&=&e^{-t}\\
v&=&2x_2e^{-t}\\
w&=&2 x_1e^{-t}\\
\end{array}
\end{equation}
It has no irregular points.\\
{\bf RTW(4) system}
\begin{equation}
\begin{array}{lll}
\dot{x}_1&=&x_3e^{-(\gamma+2)t}-2x_2^2e^{\gamma t}\\
\dot{x}_2&=&2x_1x_2e^{\gamma t}\\
\dot{x}_3&=&-2x_1x_3e^{\gamma t},\\
\end{array}
\end{equation}
where $\gamma$ is an arbitrary constant. The matrix $J$ is given
by $\Psi=(x_1^2-x_2^2)e^{\gamma t} +x_3e^{-(\gamma+2)t}$,
$\mu=1$ and the Hamiltonian is $H=x_2x_3$.\\
The reduced equations are
\begin{equation}
\begin{array}{lll}
x_1&=&\left(C_1e^{-\gamma t}- x_3e^{-2(\gamma+1)t}+\displaystyle{\frac{C_2^2}{x_3^2}}\right)^{\frac{1}{2}}\\
x_2&=&\displaystyle{\frac{C_2}{x_3}}\\
\dot{x}_3&=&-2\left(C_1e^{-\gamma t}- x_3e^{-2(\gamma+1)t}+\displaystyle{\frac{C_2^2}{x_3^2}}\right)^{\frac{1}{2}}x_3e^{\gamma t}\\
\end{array}
\end{equation}
The Poisson structure is given by
\begin{equation}
\begin{array}{lll}
u&=&e^{-(2+\gamma)t}\\
v&=&2x_2e^{\gamma t}\\
w&=&2 x_1e^{\gamma t}\\
\end{array}
\end{equation}
It has no irregular points.\\
{\bf RTW(5) system}
\begin{equation}
\begin{array}{lll}
\dot{x}_1&=&\delta x_2+x_3-2x_2^2e^{-2t}\\
\dot{x}_2&=&-\delta x_1+2x_1x_2e^{-2t}\\
\dot{x}_3&=&-2x_1x_3e^{-2t},\\
\end{array}
\end{equation}
where $\delta$ is a non-vanishing constant. The matrix $J$ is
given by $\Psi=\displaystyle{\frac{\delta
e^{-2t}}{2}}(x_1^2-x_2^2)+ \displaystyle{\frac{\delta }{2}}x_3$,
$\mu=1$ and the Hamiltonian is $H=x_1^2+x_2^2 +\displaystyle{\frac{2}{\delta }}x_2x_3$.\\
The reduced equations are
\begin{equation}
\begin{array}{lll}
x_1&=&\left(C_1e^{2t}+ x_2^2+e^{2t}\displaystyle{\frac{C_2-C_1e^{2t}-2x_2^2}{\frac{\delta}{2}x_2+e^{2t}}}\right)^{\frac{1}{2}}\\
x_3&=&\displaystyle{\frac{C_2-C_1e^{2t}-2x_2^2}{\frac{\delta}{2}x_2+e^{2t}}}\\
\dot{x}_2&=&-\delta\left(C_1e^{2t}+x_2^2+e^{2t}\displaystyle{\frac{C_2-C_1e^{2t}-2x_2^2}{\frac{\delta}{2}x_2+e^{2t}}}\right)^{\frac{1}{2}}\\
& & +2\left(C_1e^{2t}+x_2^2+e^{2t}\displaystyle{\frac{C_2-C_1e^{2t}-2x_2^2}{\frac{\delta}{2}x_2+e^{2t}}}\right)^{\frac{1}{2}}x_2e^{-2t}\\
\end{array}
\end{equation}
The Poisson structure is given by
\begin{equation}
\begin{array}{lll}
u&=&e^{-(2+\gamma)t}\\
v&=&2x_2e^{\gamma t}\\
w&=&2 x_1e^{\gamma t}\\
\end{array}
\end{equation}
It has no irregular points.

\vspace{0.3cm}

{\footnotesize
\begin{tabular}{llll}
\hline
system & \multicolumn{2}{c}{Poisson matrix}& Hamiltonian\\
& $\Psi$ & $\mu$ &\\
\hline \hline Euler &$-\frac{1}{2}(x^2_1+x^2_2+x^2_3)$&$1$
&$\frac{x^2_1}{2I_1}+\frac{x^2_2}{2I_2}+\frac{x^2_3}{2I_3}$\\
\hline Lotka-Voltera &$\ln\frac{x_3^cx_2^{bc}}{x_1}$ &$x_1x_2x_3$&
$a(bx_1-x_3)+x_2+\ln \frac{x_2^\nu}{x_3^\mu}$\\
\hline
Lorenz &$\frac{1}{4}(x_3-x_1^2)$&$1$&$x^2_2+x^2_3$\\
\hline Kermac-Mackendric &$x_1+x_2+x_3$&$x_1x_2$&
$a\ln x_1+rx_3$ \\
\hline May-Leonard
&$\frac{1}{(1-\alpha)^2}(x_2^{1-\alpha}-x_1^{1-\alpha})$&$1$&$x_1^{1-\alpha}-x_3^{1-\alpha}$\\
\hline Maxvel-Bloch &$-{\frac{1}{2\upsilon}}(x_2^2+x_3^2)$&$1$&
${\frac{1}{2}\alpha(x_2^2+x_3^2)- \frac{1}{\upsilon}(x_3+x_1^2)}$\\
\hline Lor.(1) &$-(\frac{r}{\sigma}x_1^2+
x_2^2)\frac{e^{(\sigma-1)t}}{4}-x_3^2\frac{e^{(1-3\sigma)t}}{4}$&$1$
&$x^2_1-2\sigma x_3$\\
\hline Lor.(3) &$-{\frac{1}{4}}x_1^2e^{-\sigma t}+
{\frac{\sigma}{2}}x_3e^{(\sigma-1)t}$&$1$
&$x^2_2+x_3^2$\\
\hline Lor.(5)
&${\frac{1}{4}}x_1^2e^{-t}-\frac{1}{2}x_3$&$1$&$-rx_1^2+x^2_2+x_3^2$\\
\hline Rab.(1) &${\frac{1}{8}}x_1^2 - {\frac{1}{8}}x_2^2
-{\frac{1}{4}}x_3^2e^{-2\nu t}$&$1$&$x^2_1+x_2^2-4hx_3$\\
\hline Rab.(2) &${\frac{1}{8}}x_1^2e^{-\nu t} + {\frac{1}{8}}
x_2^2e^{-\nu
t}-{\frac{1}{2}}hx_3$&$1$&$x^2_1-x_2^2-2x_3^2$\\
\hline Rab.(3)
&${\frac{1}{4}}x_2^2e^{(\nu_3-2\nu)t}+{\frac{1}{4}}x_3^2e^{-\nu_3
t}$&$1$&$x_1^2+x_2^2$\\
\hline Rab.(4) &$-{\frac{1}{4}}x_1^2e^{-\nu t} -
{\frac{1}{4}}x_2^2e^{\nu_1 t}+hx_3e^{\nu_1 t}$&$1$&$x_2^2+(h-x_3)^2$\\
\hline Rab.(5) &${\frac{1}{4}}x_1^2e^{\nu_2 t} +
{\frac{1}{4}}x_2^2e^{-\nu_2 t}-hx_3e^{\nu_2 t}$&$1$&$x_1^2-(h+x_3)^2$\\
\hline Rab.(6) &$-{\frac{1}{4}}x_1^2e^{-\nu_1 t} -
{\frac{1}{4}}x_2^2e^{(\nu_1-2\nu_2) t}$&$1$&$x_2^2+x_3^2$\\
\hline Rab.(7) &${\frac{1}{4}}x_1^2e^{(\nu_2-2\nu_3) t} +
{\frac{1}{4}}x_2^2e^{-\nu_2t}$&$1$&$x_1^2-x_3^2$\\
\hline RTW.(1)
&${\frac{1}{2}}(x_1^2-x_2^2+x_3e^{-t})$&$1$&$x_3(2x_2-\delta)$\\
\hline RTW.(2)
&$-\frac{\delta}{2}x_1^2-(\frac{\delta}{2}x_2^2+x_3x_2)e^{-t}$&$1$&$x_1^2+x_2^2+x_3$\\
\hline RTW.(3)
&$(x_1^2-x_2^2+x_3)e^{-t}$&$1$&$x_2x_3$\\
\hline RTW.(4)
&$(x_1^2-x_2^2)e^{\gamma t} +x_3e^{-(\gamma+2)t}$&$1$&$x_2x_3$\\
\hline RTW.(5) &$\displaystyle{\frac{\delta
e^{-2t}}{2}}(x_1^2-x_2^2)+ \displaystyle{\frac{\delta }{2}}x_3$
&$1$&$x_1^2+x_2^2 +\displaystyle{\frac{2}{\delta }}x_2x_3$\\
\hline \hline
\end{tabular}}

\vspace{2cm}

\noindent {\bf Table }. Examples of Hamiltonian systems given in
the text. In each example we give a Hamiltonian $H$ and a Poisson
structure $J$ ($J$ is given in terms of $\mu,\Psi$ by the
equation~(\ref{GenSol})).

\section*{4. Conclusion}

We considered the Jacobi equation for the case $N=3$. We have
found the most general Poisson structure $J$ in the neighborhood
of regular points. This form is quite suitable for the study of
the multi-Hamiltonian structure of the system. We found all
possible compatible Poisson structures and corresponding
bi-Hamiltonian systems. We studied our solution in neighborhood of
the irregular points of the Poisson structure and showed that our
it keeps its form. As an application of our results we gave
several examples which were reported earlier \cite{NG}-\cite{PR}
as bi-Hamiltonian systems. In these examples we give the Casimirs,
components of the Poisson matrix, the reduced equations and
irregular points. Among all examples, that we observed,  only the
Lotka-Voltera system has a special position. Our solution is not
valid in the neighborhood of irregular points for this system. On
the other hand the Hamiltonian function is not defined at such
points as well. Hence the Lotka-Voltera equation dose not have the
Hamiltonian formulation in the neighborhood of such points.

\vspace{1cm}

We would like to thank Dr. J. Grabowski and Dr. G. Marmo for
drawing our attention to their work \cite{GMP}. We thank to Dr P.
Olver for his constructive criticisms. We also thank to the
referee for his/her several suggestions. This work is partially
supported by the Scientific and Technical Research Council of
Turkey and by the Turkish Academy of Sciences.

\end{document}